\documentclass[preprint,12pt]{elsarticle}

\usepackage{amssymb}
\usepackage{amsmath} 

\usepackage[version=3]{mhchem} 
\usepackage{graphicx}
\usepackage{float} 
\usepackage{textcomp,mathcomp} 
\usepackage{siunitx}
\usepackage{amsmath} 
\usepackage{arydshln} 

\journal{Applied Surface Science}

\begin{document}

\begin{frontmatter}



\title{Solvent-Directed Femtosecond Laser Ablation: Tuning Phase and Defect Engineering in Hybrid CdPS$_{3}$/CdS Nanostructures}


\author[label1]{Andrei Ushkov\corref{cor1}}
\author[label1,label2]{Nadezhda Belozerova}
\author[label3]{Gleb Tikhonowski}
\author[label1]{Stepan Klimov}
\author[label1,label3,label4]{Alexander Syuy}
\author[label1]{Sergey V. Bazhenov}
\author[label1]{Sergey Novikov}
\author[label1]{Vladimir G. Leiman}
\author[label1,label3]{Aleksey Arsenin}
\author[label3]{Gleb I. Tselikov}
\author[label3]{Valentyn Volkov}

\cortext[cor1]{Corresponding author}

\affiliation[label1]{organization={Moscow Center for Advanced Studies},
            addressline={Kulakova Str. 20},
            city={Moscow},
            postcode={123592},
            country={Russia}}
\affiliation[label2]{organization={Frank Laboratory of Neutron Physics, Joint Institute for Nuclear Research},
            addressline={Joliot-Curie 6},
            city={Dubna},
            postcode={141980},
            country={Russia}}
\affiliation[label3]{organization={Emerging Technologies Research Center, XPANCEO, Internet City, Emmay Tower},
            addressline={Al Sufouh 2}, 
            city={Dubai},
            postcode={123592}, 
            country={United Arab Emirates}}
\affiliation[label4]{organization={Department of General Physics, Perm National Research Polytechnic University},
            city={Perm},
            postcode={614990},
            country={Russia}
}           
\begin{abstract}
The limited visible-light absorption of wide-bandgap van der Waals crystals fundamentally restricts their utility in solar energy conversion. Here, we report a surfactant-free, solvent-directed laser synthesis strategy to engineer the phase and optoelectronic properties of Cadmium Phosphorus Trisulfide (CdPS\textsubscript{3}). By exploiting the non-equilibrium thermodynamics of femtosecond pulsed laser ablation in liquid (fs-PLAL), we demonstrate a tunable transition from the stoichiometric ternary phase to a highly active binary-rich heterostructure. While ablation in water preserves the monoclinic CdPS\textsubscript{3} lattice, the reducing environment of isopropanol triggers the formation of CdS quantum dots and metallic cadmium defect sites. This solvent-induced phase engineering transforms the ultraviolet-active host into a robust visible-light photocatalyst. The resulting hybrid CdPS\textsubscript{3}/CdS nanocolloids exhibit superior charge separation efficiency, driven by Schottky-like metal-semiconductor junctions, achieving $\sim$ 90\% degradation of Methylene Blue under 532 nm irradiation within 30 minutes. This work establishes fs-PLAL as a scalable defect-engineering tool for complex ternary layered materials, offering a new design of high-performance metal-thiophosphate-based photocatalysts.

\end{abstract}



\begin{keyword}
Cadmium Phosphorus Trisulfide; Laser Ablation; Hybrid Nanoparicles; Photodegradation.


\end{keyword}

\end{frontmatter}



\section{Introduction}

The global efforts toward sustainable energy and environmental remediation requires the development of advanced functional materials for highly-efficient solar energy harvesting. Semiconductor photocatalysts was found to be simultaneously efficient for water purification and green hydrogen production \cite{li2018dual,li20223d,altuner2023highly}. However, conventional metal oxides (e.g., TiO\textsubscript{2}, ZnO) are constrained by wide bandgaps that limit their operation to the ultraviolet (UV) region. Consequently, the search for visible-light-active semiconductors with tunable electronic structures and high specific surface area is highly demanded, directing attention toward the family of two-dimensional (2D) transition metal phosphorus trichalcogenides (MPX\textsubscript{3}) \cite{wang2022two,samal2021two}.

Among these van der Waals (vdW) antiferromagnets, Cadmium Phosphorus Trisulfide (CdPS\textsubscript{3}) stands out for its unique structural framework, where Cd$^{2+}$ cations are coordinated by ethane-like $^{4-}$ dimers, see Fig.\ref{Article1}a. While possessing remarkable chemical stability and favorable band edge positions for redox catalysis, bulk CdPS\textsubscript{3} is an indirect wide-bandgap semiconductor ($E_g \approx 3$ eV) \cite{li2021layered,calareso1997optical,curro1998x}, which makes it practically transparent to visible light. To unlock its catalytic potential, bandgap engineering via heterostructuring (specifically coupling CdPS\textsubscript{3} with narrow-gap sensitizers like CdS) is essential. Yet, conventional "wet" chemistry synthesis routes are often slow and reliant on surface-passivating ligands that block the charge transfer \cite{rothfuss2023linker,bhavani2023surface}.

Herein, we introduce a scalable, "green," and ligand-free strategy to fabricate hybrid CdPS\textsubscript{3}/CdS nanostructures via femtosecond pulsed laser ablation in liquid (fs-PLAL). Unlike thermal equilibrium methods, fs-PLAL generates a high-density plasma plume confined within a cavitation bubble, creating extreme transient conditions that drive non-equilibrium phase transformations. Previously we have successfully synthesized  nanoparticles from the bulk vdW and specifically TMDC crystals \cite{tselikov2025tunable,tselikov2022transition,ushkov2024tungsten}. In this study, we demonstrate that solvent chemistry acts as a master switch for material design: while aqueous ablation preserves the ternary CdPS\textsubscript{3} phase, organic solvents trigger a reductive dissociation, yielding CdS quantum dots and metallic defects within the vdW matrix. This solvent-directed phase engineering results in a hybrid photocatalyst with exceptional visible-light activity toward organic pollutant degradation, establishing a versatile route for the defect-engineering of complex 2D materials.

\section{Results and discussion}
\subsection{Laser Ablation of bulk CdPS\textsubscript{3} crystal}
\label{struct_analysis}

\begin{figure}[H]
\centering\includegraphics[width =1\textwidth]{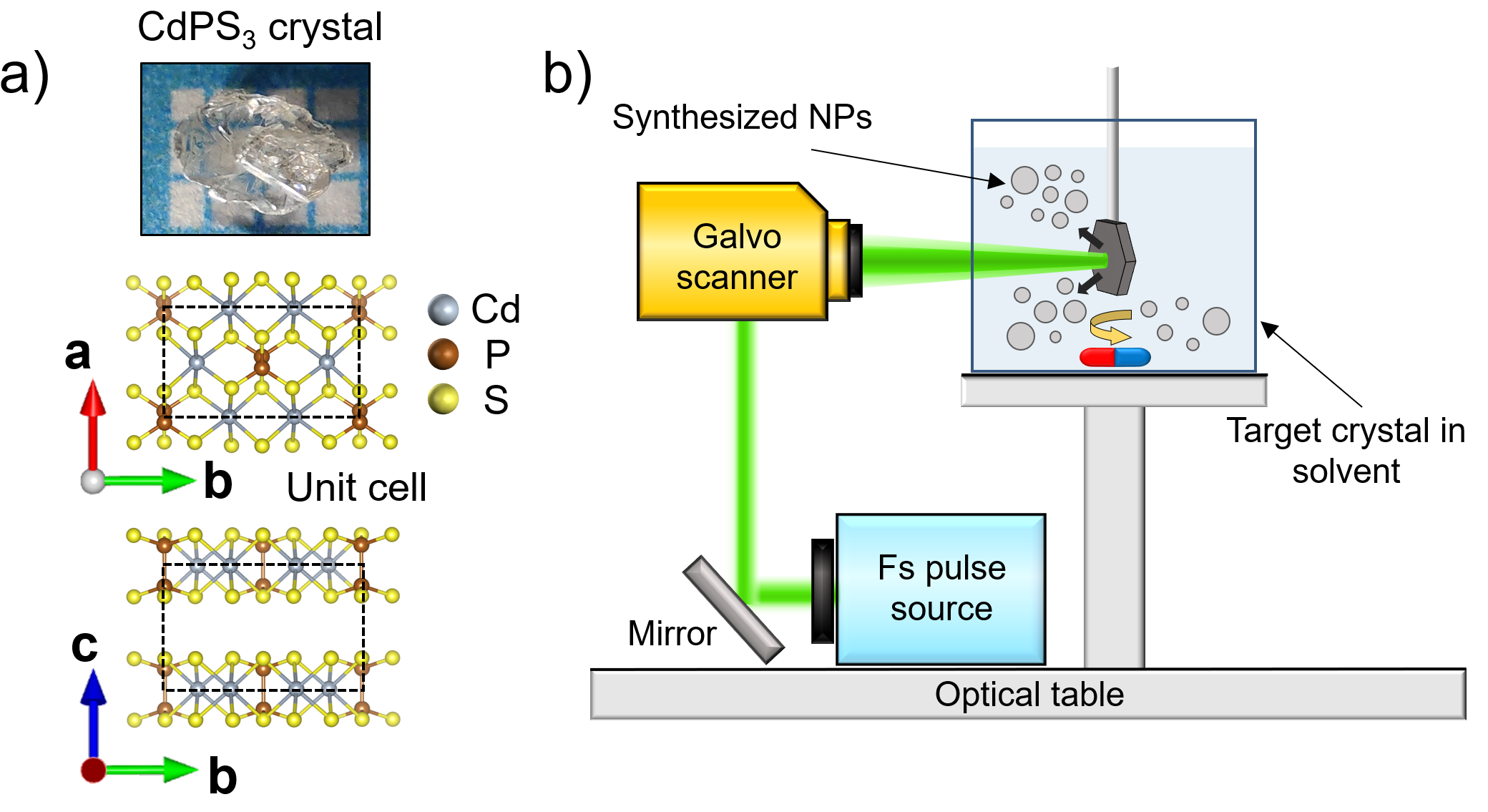}
\caption{(a) Crystal structure of bulk CdPS\textsubscript{3}; b) Scheme of experimental setup for femtosecond pulsed laser ablation in liquid.}
\label{Article1}
\end{figure}


The experimental setup is shown in Fig.\ref{Article1}b. The bulk crystalline target of CdPS\textsubscript{3} (2D Semiconductors Inc.) was fixed vertically inside a glass chamber filled with a liquid medium: deionized (DI) water, isopropyl alcohol (IPA) or acetonitrile (ACN). The NPs formation resulted in yellowish coloration of the colloidal solution, a characteristic optical signature of wide-bandgap semiconductor NPs (CdPS\textsubscript{3} and CdS), absorbing in the blue-UV region.

\subsection{Morphological And Structural Characterization of PLAL-synthesized colloids}

\begin{figure}[H]
\centering\includegraphics[width =1\textwidth]{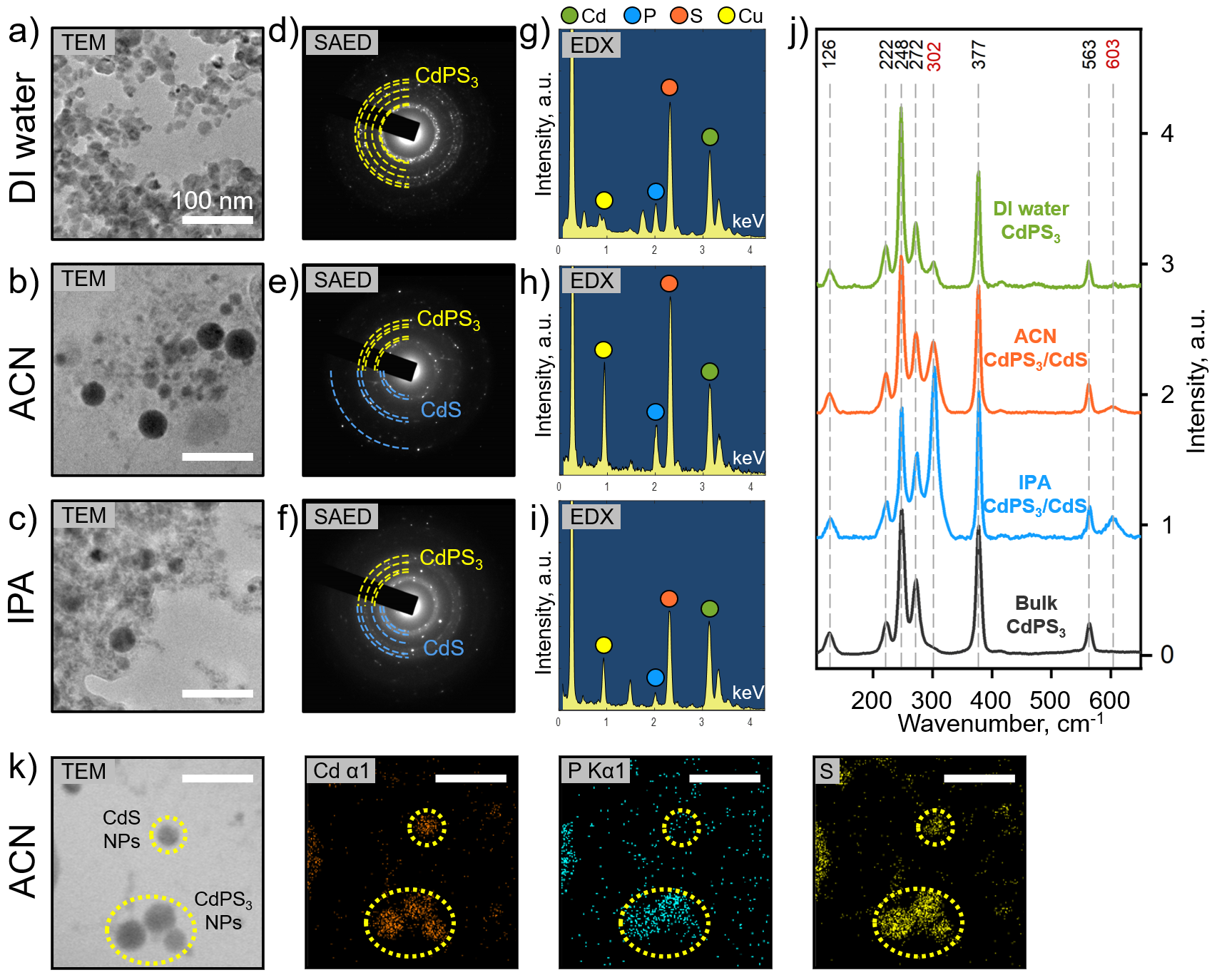}
\caption{Structural and elemental characterization of CdPS\textsubscript{3}/CdS nanoparticles synthesized in different solvents. Typical (a-c) TEM images, (d-f) corresponding SAED patterns, and (g-i) EDX spectra of CdPS\textsubscript{3} NPs synthesized in (a, d, g) deionized (DI) water, (b, e, h) acetonitrile (ACN), and (c, f, i) isopropanol (IPA). (j) Comparison of the Raman spectra of a bulk CdPS\textsubscript{3} crystal with NPs synthesized in IPA, ACN, and DI water. k) Spatially resolved elemental analysis of NPs in ACN, demonstrating distinct CdPS\textsubscript{3} and CdS NPs.}
\label{Article2}
\end{figure}

A multi-modal characterization including Transmission Electron Microscopy (TEM), Selected Area Electron Diffraction (SAED), Energy Dispersive X-ray Spectroscopy (EDX), and Raman spectroscopy was employed to elucidate the morphological, structural, and chemical properties of the colloidal nanoparticles.

Typical bright-field TEM micrographs of the colloids synthesized in DI water, ACN, and IPA are presented in Figs.\ref{Article2}a,b,c, respectively. The images reveal polydisperse colloids (NPs diameters typically range from 10 to 80 nm) in all three samples, which is intrinsic to the PLAL mechanism - the "bottom-up" nucleation and growth of spherical nanoparticles from the laser-induced plasma plume. Notably, the particles in ACN (Fig.\ref{Article2}b) and IPA (Fig.\ref{Article2}c) appear to exhibit a higher degree of sphericity compared to those in water, hinting at different surface tension and cavitation bubble dynamics in organic solvents which may favor the spherodization of products prior to solidification.

The crystallinity and phase composition were probed using SAED, with corresponding patterns shown in Figs.\ref{Article2}d,e,f. All samples display well-defined diffraction rings, confirming the crystalline nature of the nanoparticles. However, a detailed indexing of the Debye-Scherrer rings exposes a profound solvent dependency in the phase evolution.

The SAED pattern for the water-synthesized colloid (Fig.\ref{Article2}d) corresponds to the monoclinic lattice of the parent CdPS\textsubscript{3} phase (see Supplementary Note S1 for details). This indicates that water, likely due to its high thermal conductivity and heat capacity, effectively conserves the stoichiometry of the ablated ternary compound, preventing significant thermal decomposition or phase segregation. The oxidative potential of water does not appear to disrupt the thiophosphate framework under these conditions.

In stark contrast, the SAED patterns for ACN (Fig.\ref{Article2}e) and IPA (Fig.\ref{Article2}f) colloids reveal a complex superposition of diffraction signatures. Rings corresponding to the original CdPS\textsubscript{3} C2/m phase coexist with distinct reflections assignable to the hexagonal wurtzite phase of CdS. The appearance of CdS rings is the first direct structural evidence of the laser-induced phase transformation. It also supports the hypothesis, that in organic solvents the lower formation energy of CdS (-0.777 eV/atom in comparison to -0.527 eV/atom for CdPS\textsubscript{3}) drives the dissociated plasma condensation into the binary sulfide phase.

The spatial arrangement of these phases - whether they form hybrid heterostructures or exist as separate particle fractions - can be examined via spatially resolved EDX mapping. The ACN sample analysis (see Fig.\ref{Article2}k) reveals a striking binary colloidal nature. The elemental maps for Cadmium (Cd), Phosphorus (P), and Sulfur (S) show distinct particle fractions: those rich in all three elements (CdPS\textsubscript{3}) and those deficient in Phosphorus but rich in Cd and S (CdS). This spatial separation suggests that in ACN, the formation of CdS may occur within the plasma plume in parallel with formation of CdPS\textsubscript{3}.

For the DI water and IPA samples, the phases were not as spatially distinct at the resolution limit of the EDX mapping, suggesting a more dense mixing or core-shell morphology. To quantify the global phase composition, a stoichiometric partitioning analysis \cite{goldstein2017scanning} was applied to the global EDX spectra (Figs.\ref{Article2}g,h,i). Assuming the material exists primarily as stoichiometric CdPS\textsubscript{3} and CdS \cite{sendeku2024deciphering}, the molar fractions were calculated based on the atomic percentages of elements. The results are summarized in Table \ref{Table1}.

\begin{table}[h]
\centering
\renewcommand{\arraystretch}{1.5} 
\setlength{\tabcolsep}{6pt}       
\resizebox{\textwidth}{!}{
\begin{tabular}{l c ccc c ccc}
\hline\hline
PLAL solvent & & Cd, at.\% & P, at.\% & S, at.\% & & CdPS$_3$, mol. \% & CdS, mol. \% & RMSE, \% \\
\cline{1-1} \cline{3-5} \cline{7-9}

DI water & & 20.9 & 16.5 & 62.5 & & 88.0 & 12.0 & 2.3 \\
\cdashline{1-1} \cdashline{3-5} \cdashline{7-9}

ACN      & & 28.0 & 14.8 & 57.3 & & 52.6 & 47.4 & 0.07 \\
\cdashline{1-1} \cdashline{3-5} \cdashline{7-9}

IPA      & & 43.1 & 6.4  & 50.5 & & 11.3 & 88.7 & 1.4 \\

\hline\hline
\end{tabular}
}
\caption{Elemental content analysis of colloids, produced via femtosecond ablation of bulk CdPS\textsubscript{3} target in different solvents.}
\label{Table1}
\end{table}

The data in Table \ref{Table1} reveals a clear trend: the organic solvents promote the formation of CdS, with IPA being the most effective, yielding a colloid dominated by the binary phase. This is attributed to the reducing nature of IPA (a secondary alcohol), which may facilitate the abstraction of phosphorus or the stabilization of sulfur vacancies, thereby driving the equilibrium toward CdS.

Raman spectroscopy provides an additional independent confirmation of these findings. Figure \ref{Article2}j compares the spectra of bulk CdPS\textsubscript{3} with the colloidal samples. The bulk crystal displays sharp characteristic peaks at low wavenumbers (e.g., 248, 272, 377 cm$^{-1}$), corresponding to the $A_{1g}$ and $E_g$ modes of the metal-thiophosphate lattice (P$_2$S$_6^{4-}$ vibrations). The DI water sample conserves the key CdPS\textsubscript{3} modes, confirming the SAED and EDX results.

The spectra for ACN and IPA samples show a marked decrease in the intensity of the CdPS\textsubscript{3}-specific modes. However, new features emerge, particularly in the 300 cm$^{-1}$ (LO mode of CdS) and 600 cm$^{-1}$ (2LO overtone) regions. The simultaneous observation of vibrational modes from both CdPS\textsubscript{3} and CdS in these samples confirms their biphasic content. The broadening of the peaks in the colloids relative to the bulk is attributed to phonon confinement effects in the nanocrystals and lattice strain induced by the rapid solidification.

\subsection{Photoluminescence of CdPS\textsubscript{3} nanoparticles}

\begin{figure}[H]
\centering\includegraphics[width =1\textwidth]{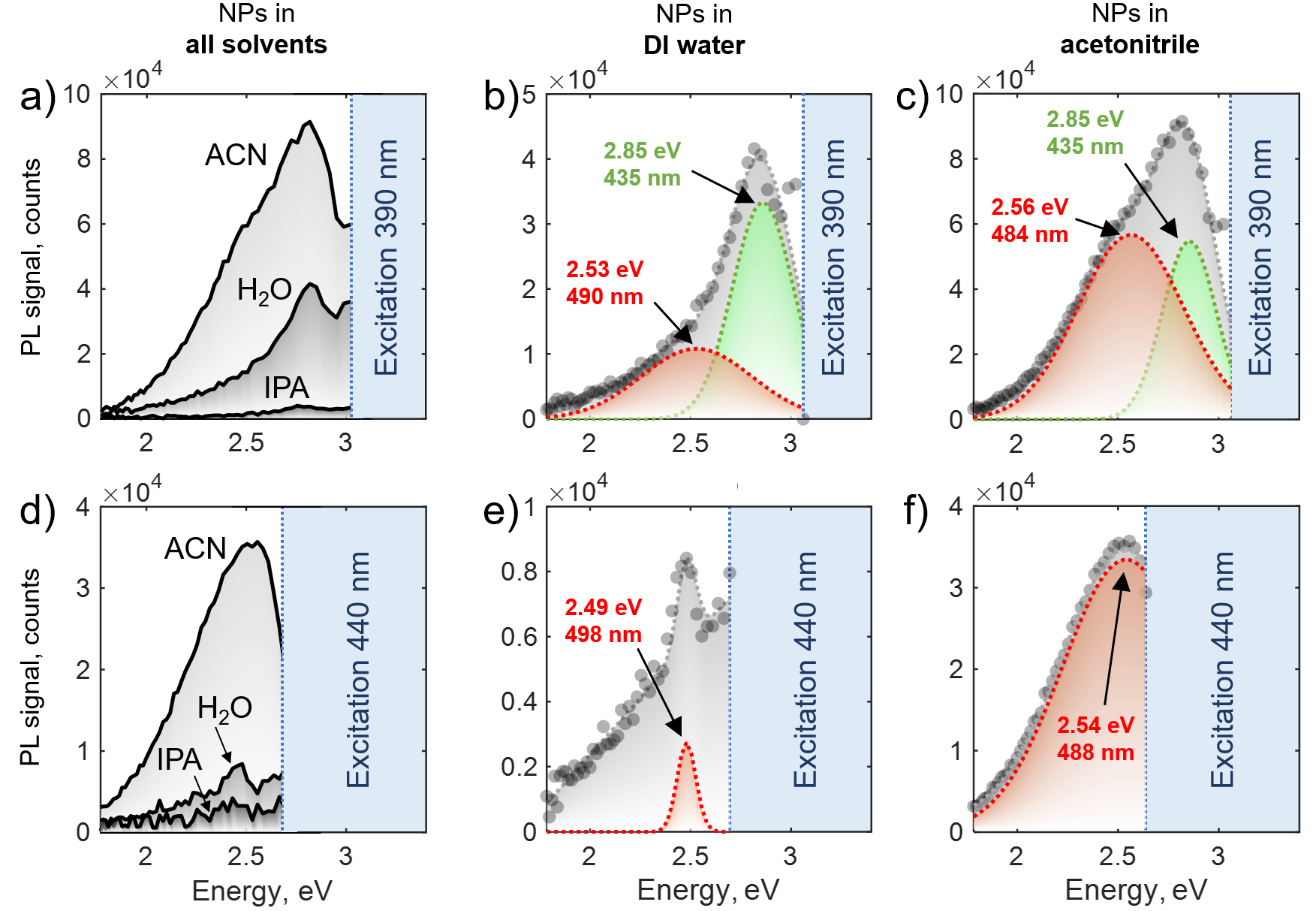}
\caption{Photoluminescence (PL) spectra of CdPS\textsubscript{3}/CdS colloids in different solvents. a) PL of ACN, DI water and IPA colloids at 390 nm excitation; b) and c) PL signal of colloidal solutions at excitation of 390 nm in DI water and ACN, correspondingly, deconvoluted into two distinct Gaussian peaks; d) PL of ACN, DI water and IPA colloids at 440 nm excitation; e) and f) PL signal of colloidal solutions at excitation of 440 nm in DI water and ACN, correspondingly, showing a distinct Gaussian peak.}
\label{Article3}
\end{figure}

To investigate the effects of solvent-mediated phase transformation on the electronic band structure, we performed a steady-state photoluminescence (PL) spectroscopy on the CdPS\textsubscript{3} and CdPS\textsubscript{3}/CdS hybrid nanoparticles.

The photoluminescence response was initially probed using an excitation wavelength of $\lambda_{exc} = 390$ nm ($3.18$ eV).1 This energy was selected to be strictly suprabandgap for the bulk CdPS\textsubscript{3} phase, which possesses an optical bandgap of approximately 3 eV ($\sim 413$ nm) \cite{li2021layered}, as well as for the binary CdS phase (bulk bandgap $\sim 2.42$ eV) \cite{sze1989physics}. The resulting emission spectra for nanoparticles synthesized in deionized (DI) water, acetonitrile (ACN), and isopropanol (IPA) are presented in Fig.\ref{Article3}a. All samples exhibited broad photoluminescence in the blue-green spectral region, yet the spectral profile varied significantly depending on the fs-PLAL solvent.

To decouple the overlapping contributions from the ternary parent phase and the laser-induced binary phase, we applied Gaussian deconvolution to the emission spectra. This analysis revealed two distinct channels, centered at approximately 435 nm ($2.85$ eV) and 490 nm ($2.53$ eV).

Figure \ref{Article3}b details the deconvolution for the DI water-synthesized colloid. The spectrum is dominated by a single emission band at 2.85 eV (435 nm). Although CdPS\textsubscript{3} is an indirect bandgap semiconductor with a low radiative efficiency, the observation of a distinct PL signal suggests the involvement of localized defect states. The non-equilibrium nature of femtosecond laser ablation, characterized by a rapid quenching rate, inevitably introduces a high density of point defects. We attribute the 2.85 eV emission to these shallow intraband defect levels, located approximately 140 meV below the conduction band edge of bulk CdPS\textsubscript{3} (3 eV). The dominance of this peak in water aligns with the structural data confirming the retention of the CdPS\textsubscript{3} stoichiometry in the aqueous environment.

In contrast, the ACN-synthesized colloid (Fig.\ref{Article3}c) exhibits a dual-emission profile with nearly equal contributions from the 2.85 eV peak and a lower-energy band centered at 2.56 eV (484 nm). This spectral equipartition provides strong optical evidence for the formation of a binary-ternary heterostructure. The emergence of the low-energy peak correlates with the appearance of CdS diffraction rings in the SAED analysis (see Fig.\ref{Article2}e). Notably, the bulk bandgap of hexagonal CdS is 2.42 eV (512 nm). We attribute this blue shift to the quantum confinement effect, indicating that the segregated CdS phase exists as quantum dots (QDs) with dimensions approaching the exciton Bohr radius of CdS ($\sim 5.8$ nm). This result suggests that the laser-induced phase transformation in ACN yields CdS nanocrystals in the quantum confinement regime, mixed with the defect-rich CdPS\textsubscript{3} NPs.

By shifting the excitation wavelength to $\lambda_{exc} = 440$ nm ($2.82$ eV), we energetically exclude the excitation of the CdPS\textsubscript{3} host (bandgap 3 eV) and its associated shallow defects (emission 2.85 eV), while remaining well above the absorption threshold of the CdS quantum dots ($\sim 2.56$ eV).As shown in Figs.\ref{Article3}e,f, this selective excitation results in the complete suppression of the high-energy peak at 435 nm. The deconvoluted spectra for both ACN and IPA samples display a single Gaussian peak centered at $\sim 490$ nm (2.53 eV). In the DI water sample (Fig.\ref{Article3}e) the signal is negligible, consisting primarily of background scattering, which further confirms the absence of a significant CdS phase in the aqueous synthesis. This control experiment conclusively demonstrates that the two PL peaks originate from chemically distinct species - CdPS\textsubscript{3} and CdS - rather than from multiple defect states within a single material phase.

A prominent PL feature is observed in the IPA-synthesized colloid. Although EDX analysis indicates the highest molar fraction of CdS ($\sim 90\%$) 1, its integrated PL intensity is significantly lower than that of the ACN or water samples (Figs.\ref{Article3}a,d). Since direct-bandgap CdS quantum dots are typically highly luminescent, this suppression implies the presence of an efficient non-radiative recombination pathway. We hypothesize the presence of metallic Cadmium (Cd$^0$) nanoparticles or clusters, reduced during ablation in the strongly reducing alcohol environment. Metallic inclusions create Schottky-like junctions at the interface with the semiconducting CdS/CdPS\textsubscript{3}, that act as electron sinks. Photo-excited electrons are rapidly transferred from the semiconductor conduction band to the metal, where they relax non-radiatively via electron-phonon scattering. While destructive to photoluminescence yield, this charge separation mechanism is highly advantageous for photocatalytic applications, as it extends the lifetime of reactive charge carriers. This hypothesis of metallic Cd active sites in IPA sample is consistent with the enhanced photodegradation efficiency of Methylene Blue, discussed in the subsequent section. Thus, the "weak" PL in IPA serves as an indirect signature of a catalytic metal-semiconductor heterojunction.

\subsection{Photodegradation properties of CdPS\textsubscript{3}/CdS nanoparticles}

\begin{figure}[H]
\centering\includegraphics[width =1\textwidth]{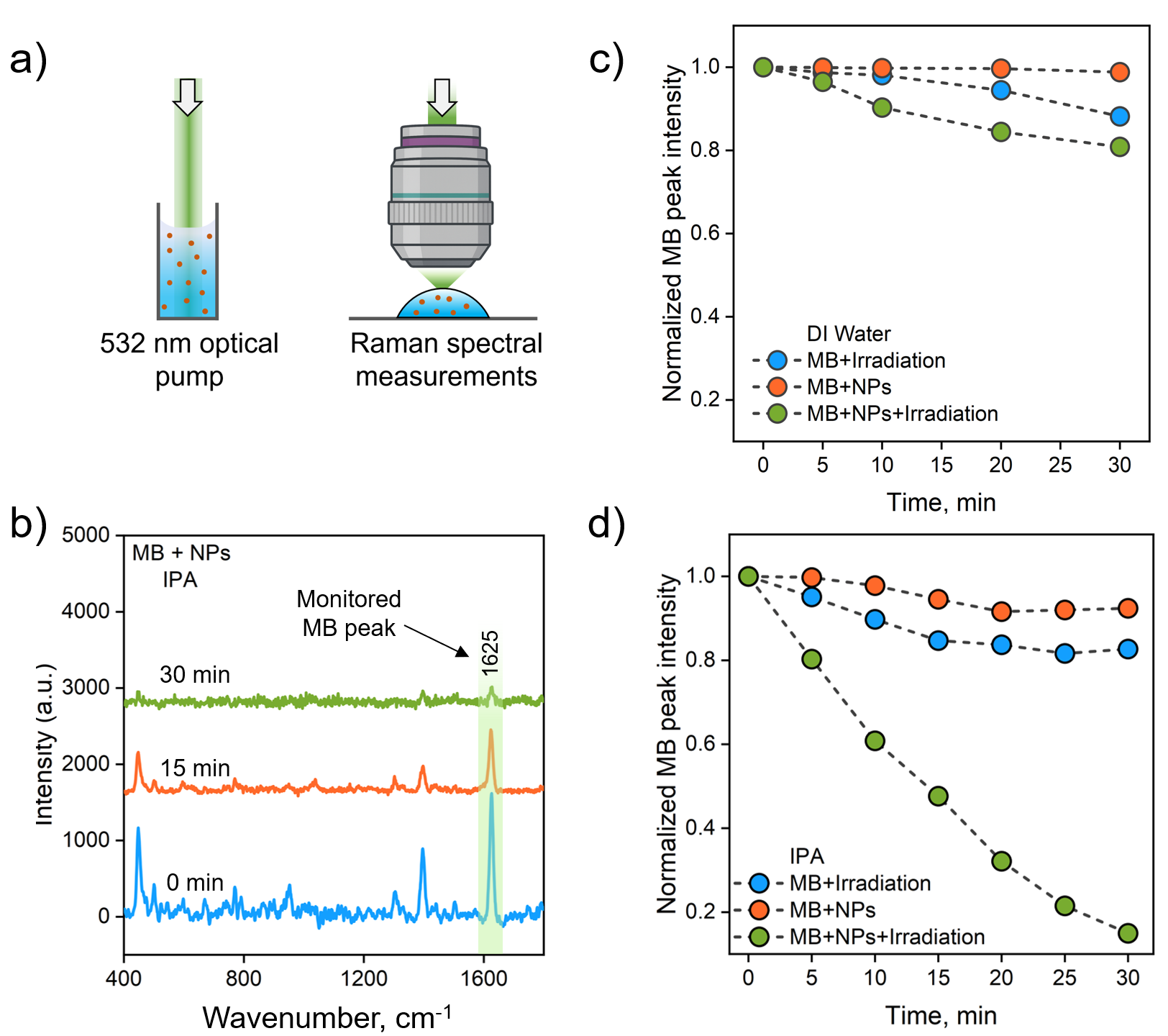}
\caption{(a) Schematic of the experimental setup. (b) Time-resolved Raman spectra of the photocatalytic system containing a 1:1 mixture of MB in isopropanol and a colloidal suspension of CdPS\textsubscript{3} nanoparticles. (c, d) Intensity of the MB Raman peak at 1625 $cm^{-1}$ as a function of time for samples in different solutions: (c) DI water and (d) IPA.} 
\label{Article4}
\end{figure}

The photocatalytic efficiency of the laser-synthesized nanoparticles was evaluated by monitoring the degradation of Methylene Blue (MB) dye, a model cationic organic pollutant, under visible light irradiation ($\lambda = 532$ nm). The kinetics of the degradation process were probed via in situ Raman spectroscopy, a technique chosen for its ability to provide molecular-level fingerprints of the pollutant concentration in real-time while distinguishing between adsorption and catalytic decomposition events. The experimental configuration, depicted in Fig.~\ref{Article4}a, utilized a confocal Raman microscope where the excitation laser served the dual function of the actinic source for driving the photocatalytic reaction and the probe beam for acquiring the Raman scattering signal of the analyte.

To rigorously decouple the contributions of photocatalysis, physical adsorption, and direct photolysis, the study employed a comparative experimental design with three distinct sample groups for each solvent system (deionized water and IPA): (1) the Photocatalytic System (MB + NPs + Irradiation), (2) the Adsorption Control (MB + NPs in dark), and (3) the Photolysis Control (MB + Solvent + Irradiation). The degradation kinetics were quantified by tracking the intensity of the characteristic vibrational mode of the phenothiazine ring of MB at approximately 1625 cm$^{-1}$, which corresponds to the $\nu$(C–C) and $\nu$(C–N) symmetric stretching vibrations \cite{rubim2001raman, vu2020sensitive}.

Fig.~\ref{Article4}b illustrates the time-resolved Raman spectral evolution of the MB solution containing nanoparticles synthesized in IPA. A pronounced and monotonic attenuation of the MB Raman signatures, particularly the 1625 cm$^{-1}$ band, was observed over the 30-minute irradiation period. The nearly complete disappearance of the signal indicates the rupture of the aromatic chromophore and effective mineralization of the dye. Conversely, the system containing nanoparticles synthesized in DI water exhibited negligible spectral changes under identical irradiation conditions, suggesting a lack of photoactivity in the visible spectral range.

The quantitative kinetic profiles for all sample groups are presented in Fig.~\ref{Article4}c (DI water) and Fig.~\ref{Article4}d (IPA). The data reveals a dramatic difference in photocatalytic performance. In the aqueous synthesis environment, the nanoparticles - previously identified as the stoichiometric CdPS\textsubscript{3} phase - showed no catalytic activity ($C/C_0 \approx 1$). This behavior is consistent with the electronic structure of bulk CdPS\textsubscript{3}, which possesses an indirect bandgap \cite{vu2020sensitive}. In sharp contrast, the IPA-synthesized colloids demonstrated rapid degradation kinetics, with the normalized Raman intensity decreasing by approximately 90\% within 30 minutes (Fig.~\ref{Article4}d). The control experiments for this system confirmed that physical adsorption (dark) and direct laser photolysis accounted for less than 10\% of the dye removal (Figs.~\ref{Article4}c, d), attributing the majority of the degradation to the photocatalytic mechanism, consistent with previous reports on MB stability under visible light in the absence of catalysts \cite{rauf2009fundamental}. 
This enhanced visible-light activity is directly correlated with the fs laser-induced phase transformation in the IPA medium. As evidenced by the structural characterization (SAED and EDX) and Raman analysis, ablation in IPA favors the formation of a CdS-rich phase. Cadmium sulfide, a direct bandgap semiconductor \cite{sze1989physics}, effectively absorbs visible radiation, generating charge carriers \cite{cheng2018cds, nasir2020recent, zhang2023research}.

Beyond this, the non-equilibrium fs-PLAL synthesis may introduce catalytic defects. The reducing IPA medium, acting as a radical scavenger\cite{pitj}, could facilitate partial reduction of Cd$^{2+}$ to metallic Cd$^{0}$ domains that function as Schottky barriers at CdS surfaces \cite{li2024situ}. These defects would synergize with the type-II CdPS\textsubscript{3}/CdS heterojunction, creating a hierarchical charge separation architecture where Cd$^{0}$ suppresses local recombination while CdPS\textsubscript{3} mediates interfacial electron transport. The combination of phase engineering (formation of visible-light-active CdS) and defect engineering (metallic Cd sites) renders the IPA-synthesized nanoparticles highly efficient for solar-spectrum remediation applications.

\section{Conclusion}
We have successfully demonstrated a surfactant-free, solvent-directed laser synthesis strategy to engineer the phase and optoelectronic properties of Cadmium Phosphorus Trisulfide (CdPS\textsubscript{3}). By exploiting the non-equilibrium thermodynamics of femtosecond pulsed laser ablation in liquid (fs-PLAL), we observed a tunable transition from the stoichiometric ternary phase to a highly active binary-rich heterostructure. While ablation in water preserves the monoclinic CdPS\textsubscript{3} lattice, the reducing environment of isopropanol triggers a reductive dissociation, yielding CdS quantum dots and metallic cadmium defect sites within the van der Waals matrix. This solvent-induced phase engineering transforms the ultraviolet-active host into a robust visible-light photocatalyst. The resulting hybrid CdPS\textsubscript{3}/CdS nanocolloids exhibit superior charge separation efficiency, driven by Schottky-like metal-semiconductor junctions, achieving $\sim$90\% degradation of Methylene Blue under 532 nm irradiation within 30 minutes. This work establishes fs-PLAL as a scalable defect-engineering tool for complex ternary layered materials, offering a new design paradigm for high-performance metal-thiophosphate-based photocatalysts in solar energy conversion and environmental remediation applications.

\section{Experimental Section}

\subsection{Nanoparticles synthesis via femtosecond laser ablation}
\label{label_synthesis}
Colloidal samples were produced using femtosecond pulsed laser ablation in liquid (fs-PLAL), with a bulk CdPS\textsubscript{3} crystal as the target. In this technique, intense laser pulses generate a plasma plume from the target, which is then trapped inside a cavitation bubble. The dynamics of this bubble's growth and implosion are critically influenced by the physicochemical properties of the liquid medium (e.g., density, heat capacity, thermal conductivity). These dynamics define the plasma condensation kinetics, which directly influence the final nanoparticle content and crystalline structure.

A $\varnothing$2 mm beam from Yb:KGW system (1030 nm, 400 fs, 10 $\mu$J, 200 kHz, Satsuma X-28 model, Amplitude, France) was used as a source of laser radiation. The bulk crystalline target of CdPS\textsubscript{3} (dimensions: 3 $\times$ 4 $\times$ 2 mm, 2D Semiconductors Inc., Phoenix AZ, USA) was fixed vertically inside a glass chamber (BK-7, wall thickness 3 mm) filled with 5 ml of liquid media (deionized water, isopropyl alcohol, acetonitrile). Laser beam was focused on the target surface by F-Theta lens (100 mm focal distance, Thorlabs, USA) with a spot diameter $\sim$ 50 $\mu$m, corresponding to a fluence of $\sim$ 0.5 J/$cm^2$. To improve the synthesis productivity the liquid thickness between the target and chamber wall was minimized down to 3 mm. To avoid the ablation of a single surface spot the laser beam was continuously moved over a 2 $\times$ 2 mm area on the target surface with 3 m/s speed by galvanometric scanner (2-Axis VantagePro, Thorlabs, USA). The duration of laser ablation was 10 minutes. The NPs formation was visually detected by yellowish coloration of the initially transparent solution.
The colloids remained stable after the laser ablation, showing no discernible change in color for at least one month. For storage purposes, samples were kept in closed 2 mL Eppendorf microcentrifuge tubes at a temperature of 6\textcelsius{} to mitigate evaporation and avoid contamination.

\subsection{Sample characterization}
\subsubsection{Electron microscopy}

Structural and morphological characterization was conducted using a high-resolution TEM (JEM 2010; JEOL) at 200 kV, utilizing a Gatan Multiscan CCD camera for both imaging (TEM imaging) and diffraction (SAED pictures) observations.

Elemental analysis and atomic composition (EDX) were assessed using a scanning TEM system (MAIA 3; Tescan) equipped with an X-act EDS detector (Oxford Instruments).

Specimens for the microscopic analysis were prepared  by depositing 2 $\mu$L of the nanoparticle suspension onto a purified silicon substrate and allowing it to dry at room temperature.

\subsubsection{Raman spectroscopy}
Raman spectroscopy was performed using a Horiba LabRAM HR Evolution system equipped with excitation wavelengths of 532 nm, a diffraction grating of 600 lines/mm, and a 100$\times$ microscope objective with a numerical aperture (NA) of 0.9. The measurements demonstrated good spectral reproducibility.

\subsubsection{Photoluminescence spectroscopy}

Photoluminescence spectroscopy was performed on colloidal suspensions of CdPS\textsubscript{3} nanoparticles using a BioTek Synergy H4 Hybrid Multi-Mode Microplate Reader (BioTek Instruments, Inc., Vermont, USA). Spectra were acquired at room temperature under controlled excitation conditions.

\subsubsection{Photodegradation Studies}
 
The photocatalytic performance of the synthesized CdPS\textsubscript{3} nanoparticles was assessed by tracking the degradation of methylene blue (MB) dye under 532 nm laser irradiation from a continuous-wave Nd:YAG source. To decouple the contributions of photocatalytic degradation, adsorptive loss, and direct photolysis, a comparative study was conducted across three distinct sample groups prepared in both isopropyl alcohol (IPA) and deionized (DI) water. The first group, designated the photocatalytic system, consisted of a 1 $\mu$M MB solution mixed in a 1:1 volume ratio with the CdPS\textsubscript{3} NPs colloid and was subjected to laser irradiation. The second group, the adsorption system, was prepared identically but was kept in the dark without laser exposure to quantify any dye removal due solely to adsorption onto the nanoparticle surfaces. The third group served as an MB reference control, comprising a 1:1 mixture of the MB solution and pure solvent (IPA or DI water), which was irradiated to account for direct photolytic decomposition of the dye. All reactions were monitored over a 30-minute period by periodically acquiring Raman spectra with a Horiba LabRAM HR Evolution spectrometer, allowing the degradation progress to be tracked via the temporal decay of characteristic MB Raman fingerprints.

\section*{Acknowledgments}

\section*{Funding}
This work was funded by the RSF, project No. 25-19-00326. Maintenance of the photoluminescence spectroscopy platform was supported by the Ministry of Science and Higher Education (agreement 075-03-2025-662, project FSMG-2025-0003).

\section*{Conflicts of Interest}
The authors declare no conflict of interest.

\section*{Author Contributions}
A.A. supervised the project. A.U. designed the experiments and wrote the manuscript. N.B. conducted the characterizations, performed the experiments and wrote the manuscript. G.T. performed the colloidal synthesis. A.S. and S.K. conducted the TEM/SAED characterizations. S.V.B. provided experimental resources. S.N., V.G.L., G.I.T., A.V.K. and V.V. provided experimental resources and edited the manuscript. All authors discussed the results. All authors have read and agreed to the published version of the manuscript.

\section*{Data Availability Statement}
The data that support the findings of this study are available from the corresponding author upon reasonable request.

\section*{Supporting Information}

\section*{Abbreviations}
The following abbreviations are used in this manuscript:\\

\noindent 
\begin{tabular}{@{}ll}
vdW & Van der Waals\\
ACN & Acetonitrile\\
DI water & Deionized water\\
IPA & Isopropyl Alcohol\\
NP & Nanoparticle\\
fs-PLAL & Femtosecond  Pulsed Laser Ablation in Liquid\\
TEM & Transmission Electron Microscope\\
SAED & Selected Area Electron Diffraction\\
EDX & Energy-dispersive X-ray spectroscopy\\
PL & Photoluminescence\\
MB & Methylene Blue\\
NA & Numerical Aperture\\
\end{tabular}

\bibliographystyle{elsarticle-num}
\bibliography{references}

\end{document}